\documentclass[preprint,prl,aps,showpacs]{revtex4}
\usepackage{axodraw}

\begin{document}

%\draft
%%%%%End of Preamble
%%%%Start of Text%%%%%%%%%%%%%%%%%%%%%%%%%%%%%%%%%%%%%%%%%%%%%%%%%%%%%%%
%\tightenlines
%\preprint{ \vbox{\halign{&##\hfil\cr & AS-ITP-00-13 \cr & \today
%\cr &\vspace{0.6truein}   \cr }}}
%\baselineskip=14pt

\title{ Describing relativistic fermions with  two-component field}

\author{Yu-Qi Chen}
\affiliation{Institute of Theoretical Physics, Academia Sinica,
Beijing 100080 }

\begin{abstract}
With a non-unitary transformation, the Lagrangian of a Dirac
fermion is decomposed into two decoupled sectors. We propose to
describe massive relativistic fermions in gauge theories in a
two-component form. All relations between the Green's functions in
this form and those in the conventional four-component form are
derived. It is shown that the $S-$matrix elements in both forms
are exactly the same. The description of the fermion in the new
form simplifies significantly the $\gamma$-matrix algebra in the
four-component form. In particular, in perturbative calculations
the propagator of the fermion is a scalar function. The advantages
of the two-component form make it very useful in practical
applications.
\end{abstract}


\maketitle

%\vfill \eject

%\narrowtext

Describing relativistic fermions using four-component spinors is
one of the most fantastic inventions made by Dirac\cite{dirac}. In
this description, a fermion and its anti-particle  are put in the
same four dimensional spinor representation of the Lorentz group.
The motion of the fermion satisfies the Dirac equation, and the
Lagrangian expressed in this field keeps manifest Lorentz
invariance and the local interaction terms. The whole formalism is
simple and elegant. However, it is not very convenient in
practical applications. For instance, in evaluating Feynman
diagrams of multi-particle processes, handling the $\gamma$-matrix
algebra turns out to be very hard and tedious, even by virtue of
modern computers. In solving bound state problems, the complicated
spinor structure of the wave function makes solutions difficult to
gain. In lattice simulations, the discretized Lagrangian of chiral
fermions suffers from the doubling problem. In order to overcome
all these drawbacks, it is crucial to develop new methods to
describe relativistic fermions.

In this paper, we propose a two-component form to describe massive
relativistic fermions. We establish the formalism in the following
way. First with a non-unitary transformation, we show that the
Lagrangian of the fermion is decomposed into two decoupled
sectors, and each of them depends on a two-component field. One of
them is in a nonlocal form similar to the unexpanded Lagrangian in
the Heavy Quark Effective Theory (HQET), while the other one is
trivial. We then derive relationships between the Green's
functions in this two-component form and those in the conventional
four-component form. Finally, with one of these relationships, we
show that the $S-$matrix elements in both forms are exactly the
same. The conclusion holds not only for fermion scattering
processes but also for fermion antifermion annihilation processes.
In this way, we show that a relativistic massive fermion and
antifermion can be uniformly described with a two-component field.

In history, a two-component field was proposed to describe a
nonrelativistic fermion by Pauli. In nonrelativistic limit the
fermion and the antifermion near on-mass shell are decoupled so
that they can be described by separate two-component fields. This
strategy has been taken in constructing HQET\cite{HQET} and
nonrelativistic QCD ( NRQCD )\cite{NRQCD}. In extreme-relativistic
case, the left hand and the right hand fermions are decoupled and
each of them can be described by a two-component field. At first
sight, it seems surprising to describe a massive relativistic
fermion and anti-fermion simultaneously with a single
two-component field. However, it is not difficult to see from the
following facts: The generic $S-$matrix elements for a fermion
scattering or fermion-antifermion annihilation processes can be
written as a bilinear form of the Dirac spinors with $4\times 4$
matrixes in between. The free four-component wavefunction
satisfying the Dirac equation can be expressed by the Pauli
two-component wavefunction. Thus eventually the $S-$matrix
elements can be reduced to a bilinear form of the Pauli spinors
with $2\times 2$ matrixes in between. In the two-component form,
we calculate the $S-$matrix elements directly from the
two-component Lagrangian and the wavefunctions from the beginning.
Here the fermion and the antifermion are distinguished via
different wavefuntions.

The description of the fermions in this form has a number of
advantages. An apparent one is that it significantly simplifies
the $\gamma-$matrix algebra. Instead of the $4\times 4 $ $\gamma$
matrixes,  only the $2\times 2 $ $\sigma$ matrixes appear in the
Lagrangian of the two-component form. Moreover, in perturbative
calculations, the propagator of the free fermion is a scalar and
the interaction vertices can be expressed as a $2\times 2 $
matrix. It is very convenient to carry out numerical calculations
for complicated high energy processes.

We start the derivation from the following observation. For a
generic Hermitan matrix $H$,
\begin{equation}
 H\;=\;
 \left(\;
\begin{array}{cc}
 A  &  B  \\
 B^\dagger & C
 \
 \end{array}\;\right)
 \;,
 \label{H}
\end{equation}
there exists a non-unitary transformation expressed by a triangle
matrix $S$,
\begin{equation}
 S\;=\;
 \left(\;
\begin{array}{cc}
 \mathrm{I}  &  0  \\
 ~~- C^{-1}\, B^\dagger ~~& \mathrm{I}
 \
 \end{array}\;\right)
 \;,
 \label{S}
\end{equation}
which makes $S^\dagger H S$ diagonalized. Specifically, we have:
\begin{equation}
S^\dagger \,H\, S \; =\; \left(\;
\begin{array}{cc}
 A - B C^{-1} B^\dagger   &  0  \\
 0 & C \
 \end{array}\;\right)
\;.
 \label{SAS}
\end{equation}

In a gauge theory, the Lagrangian density of a fermion is given by
\begin{eqnarray}
{\cal L} \;&=&\; \overline \Psi(x) \, ( i{\not\!
D}\,-\,m\,)\,\Psi(x)\;, \label{L}
\end{eqnarray}
where $D_\mu=\partial_\mu - i g A_\mu  $ is the covariant
derivative. To make the transformation on the Lagrangian density,
we follow the technique adopted in HQET and decompose the Dirac
four-component field $\Psi(x)$ as\cite{HQET}:
\begin{eqnarray}
\Psi(x) \;&=&\; \Psi_{+}(x) \,+\, \Psi_{-}(x)\;,
 \label{psi}
\end{eqnarray}
where $\Psi_{\pm}(x) =\mathrm{P}_{\pm}\,\Psi(x) $ and the
projection operators $\mathrm{P}_{\pm}$ are defined by
\begin{equation}
\mathrm{P}_{\pm} \;\equiv \; \frac{1\,\pm\, \not\!{V}}{2}\;,
 \label{ppm}
\end{equation}
with the four-velocity parameter $V$ satisfying $V^2=1$. The
projection operators divide the four dimensional spinor space into
two two-dimensional spinor subspaces. With this decomposition, the
matrix $L \equiv i\,{\not\! D}\,-\,m\,$ can be rewritten as:
\begin{eqnarray}
 L \;&=& \;
 \left(\;\begin{array}{cc}
   iD\cdot V -m~~ &  i{ D}\cdot\sigma_\perp  \\
   i{ D}\cdot\bar{\sigma}_\perp  ~~& -iD\cdot V-m \
 \end{array}\;\right)\;,
 \label{Lv}
\end{eqnarray}
where $\sigma^\mu  \equiv(1,\mbox{\boldmath $\sigma$}) $ and
$\bar{\sigma}^\mu \equiv (1,-\mbox{\boldmath $\sigma$}) $;
 $ \sigma_\perp^\mu \equiv \sigma^\mu - \sigma \cdot V \,V^\mu$
and $ \bar{\sigma}_\perp^\mu \equiv \bar{\sigma}^\mu -
\bar{\sigma} \cdot V \,V^\mu$.

According to Eq.~(\ref{S}), choosing the transformation matrix $S$
as:
\begin{equation}
 S\;=\;
 \left(\;
\begin{array}{cc}
 \mathrm{I}   &  0  \\
 ~\displaystyle {1\over iD\cdot V +m} \,
  i{ D} \cdot \bar{\sigma}_\perp   ~~& \mathrm{I}  \
 \end{array}\;\right)
 \;,
 \label{Sv}
\end{equation}
and defining $\bar{S}\equiv \gamma^0 \, S^\dagger \,\gamma^0$,
$L$ transforms into:
\begin{eqnarray}
L' \;&=&\; \bar{S} \,L\,S \nonumber \\
 \;&=&\;
 \left(\;\begin{array}{cc}
   iD\cdot V -m\,-\, { D} \cdot \sigma_\perp\;
   \displaystyle{1 \over iD\cdot V+m}\;
   { D} \cdot \bar{\sigma}_\perp & 0 \\
   ~~ 0 & ~~-iD\cdot V-m
 \end{array}\;\right) \;.
 \label{L-D}
\end{eqnarray}
Correspondingly, the field $\Psi(x)$ transforms into $\Psi'(x)=
S^{-1} \; \Psi(x)$.
 Its component form reads:
\begin{eqnarray}
 \left(\;\begin{array}{c}
   \Psi'_{+}(x) \\
   \Psi'_{-}(x)
 \end{array}\;\right) \;&=&\;
  \left(\;\begin{array}{c}
   \Psi_{+}(x) \\
   \Psi_{-}(x)\,-\,\Psi^c_{-}(x)\,
 \end{array}\;\right)\; ,
 \label{psi'}
\end{eqnarray}
where
\begin{equation}\label{psic}
  \Psi^c_{-}(x) \;=\;  \displaystyle{ 1 \over iD \cdot V+m}\;
  i{ D} \cdot \bar{\sigma}_\perp \; \Psi_{+}(x)\; .
\end{equation}
We see that $\Psi'_-(x)=0$ if $\Psi_{-}(x)$ satisfies  the Dirac
equation.

The Lagrangian density of the fermion can then be written in a
compact form as:
\begin{eqnarray}
{\cal L} \;&=&\; \overline{\Psi}'(x)  \;L' \;
   \Psi'(x)
 \;.
 \label{L'}
\end{eqnarray}

Eqs.~(\ref{L-D})-(\ref{L'}) imply that after the transformation,
the fermion Lagrangian density is decomposed into two independent
sectors. They are decoupled, and each of them is described by a
two-component field. It is interesting that the first sector is
similar with the nonlocal form  of HQET which was derived using
the equation of motion or the generating functional method. In
these approaches, certain approximations have been made. Here
there is no approximation being made when we derive it via the
non-unitary transformation (\ref{Sv}). This ensures the
equivalence of the four-component form and the two-component form.

The first term is invariant under the reparameterization
transformation\cite{Chen:1993sx}. This ensures that the physical
predictions made from it are independent of the choice of the
velocity parameter $V$ in spite of an explicit $V$-dependence in
the decomposition (\ref{psi}).
%Later on we will show that the
%$S-$matrix element with fermions in the external lines can be
%extracted from the nonlocal term.
%This implies that the original
%four component form can be equivalently described by the two
%component form.

The original Lagrangian density (\ref{L}) can be recovered by the
inverse transformation:
\begin{eqnarray}
 L&=& \bar{S}^{-1}\, L'\, {S}^{-1} \;, ~~~~
 \Psi(x)\;=\; {S} \, \Psi'(x) \;.
 \label{S-inv}
\end{eqnarray}
The transformation leads to the following field identity:
\begin{equation}\label{Gv-G}
\Psi(x) \; \overline{\Psi} (y)\; = \; S (x)\;\Psi'(x) \;
\overline{\Psi}'(y)\;\bar{S} (y)\;.
\end{equation}
This identity implies a set of relations between the Green's
functions in both forms. With $S(x)$ given in Eq.(\ref{Sv}), it
follows that:
\begin{eqnarray}
  \mathrm{P}_{+} \,\langle\, 0\, |\, T\,\Psi(x)\,
  \overline{\Psi}(y)\,
  |\, 0\,\rangle^B \,\mathrm{P}_{+}\,
 \;&=&\;
 \langle\, 0\, |\,T\, \Psi'_{+}(x)\,\overline{\Psi}'_{+}(y)\,
  |\, 0\, \rangle^B
\label{ppp} \,, \\
 \mathrm{P}_{-} \,\langle\, 0\, | \,T\,\Psi(x)\,\overline{\Psi}(y)\,|\, 0\, \rangle^B\,
 \mathrm{P}_{+}
 \;&=&\;
 \langle \,0\, |\,T\, \Psi^c_{-}(x)\,
 \overline{\Psi}'_{+}(y)\,|\, 0\, \rangle^B
 \label{pmp} \;, \\
 \mathrm{P}_{-} \,\langle\, 0\, | \,T\,\Psi(x)\,
 \overline{\Psi}(y)\,|\, 0\, \rangle^B\,
 \mathrm{P}_{-}
 \;&=&\;
 \langle \,0\, |\,T\,
\Psi'_{-}(x)\,\overline{\Psi}'_{-}(y)\,|\, 0\, \rangle^B
 \nonumber \\
 \;&-& \;
\langle \,0\, |\,T\, \Psi^c_{-}(x)\,
\overline{\Psi}^c_{-}(y)\,|\,
0\, \rangle^B
 \label{pmm} \;,
\end{eqnarray}
where we have omitted the  conjugation relation of (\ref{pmp}).
The superscript $B$ denotes that they are the Green's functions in
an arbitrary background field. With the path integral method, any
Green's function with gluon external lines can be obtained by
functional derivative with respect to the background field $B(x)$.
Since the determinant of the transform (\ref{Sv}) is unity, the
measure of the integration element of the fermion field remains
unchanged. This ensures the validation of these relations beyond
the tree level. From these relations we see that once the right
hand side of each relation is calculated in the two-component
form, all possible projected Green's functions in the original
theory can be obtained, and {\it vice versa}. This implies that
both forms are equivalent. Thus these relations signify
connections between the four-component form and the two-component
one.

In HQET, relation (\ref{ppp}) has been used as matching
conditions\cite{Grinstein:1990mj} to determine the renormalized
effective Lagrangian . Relation (\ref{pmp}) has been used as
matching conditions\cite{Chen:2002yr} to determine the
renormalized transformation of the heavy quark field in the
reparametrization invariance. The relation (\ref{pmm}) is a new
one, which has not yet been derived before.

 We now show that the $S-$matrix elements evaluated with both
forms are the same. For a process of a fermion scattered by
arbitrary gluons or fermion anti-fermion annihilation or creation,
a generic form of the Green's functions in the momentum space can
be expressed as:
\begin{equation}
G(p ,p' ;\Gamma )\;  = \; G( p )\;\Gamma \;G( p' ) \;,
 \label{G-Gamma}
\end{equation}
where $p$ and $p'$ are the external momenta of the fermions;
$G(p)$ and $G(p')$ are the propagators of the free fermions;
$\Gamma$ denotes the interaction vertex. Using the reduction
formula, the $S-$matrix element for the fermion scattering process
in the four-component form can be obtained as follows:
\begin{eqnarray}
 T_{s\,s'}(p'\to p + X ) \;& = &\;
 \lim\limits_{p^2,p'^2\to m^2}\; \bar u_s (p )
 \;G^{ - 1} (p )\;G(p ,p' ;\Gamma )\;G^{ - 1}(p' )\;u_{s'}(p' )
 \nonumber \\
 \; &=&\;  \bar u_{s} (p )\;\Gamma \;u_{s'}(p' ) \;,
 \label{T-scat}
\end{eqnarray}
where $u_{s} (p)$ and $u_{s'}(p' )$ are the four-component Dirac
wave functions of the fermions. Similarly, the $S-$matrix element
for the fermion anti-fermion annihilation process reads
\begin{eqnarray}
 T_{s\,s'}(\bar{p} + p' \to X )
 \;& = &\;
 \lim\limits_{p^2,p'^2\to m^2}\; \bar v_s (p )
 \;G^{ - 1} (-p )\;G(-p ,p' ;\Gamma )\;G^{ - 1}(p' )\;u_{s'}(p' )
 \nonumber \\
 \;  &=&
 \;\bar v_{s} (p )\;\Gamma \;u_{s'}(p' ) \;,
\label{T-anni}
\end{eqnarray}
where  $v_{s}(p)$ is the Dirac wavefunction of the antifermion.

To relate the $S-$matrix in the four-component form to that in the
two-component form, we first write down the wavefunction of the
fermion in the two-component description. For simplicity, we take
the four-velocity parameter $V=(1,0,0,0)$. However, all results
can easily be generalized to an arbitrary $V$.
\begin{eqnarray}
\tilde u_s(p) &\; = \;& \mathrm{P}_{+} u_s(p)\; = \; {\sqrt {E +
m} }
 \;\;{\mathbf{\xi }_s} \label{tilde-u}\\
\tilde v_s(p) &\; = \;& \mathrm{P}_{+} v_s(p)\; = \;
\frac{{\bf{p}} \cdot \mbox{\boldmath $\sigma$} }
 {\sqrt {E+ m }}\;\;{{\xi }_s}\; ,
  \label{tilde-v}
\end{eqnarray}
where $E\equiv p_0$, and $\xi_s$ ($s=1,2$) are the Pauli
two-component spinors.

% :
%
%$ \xi_1 \;=\; \left( \; \begin{array}{c}
%  1  \\
%  0  \\
%\end{array} \; \right)$,
%$\xi_2 \;=\; \left( \; \begin{array}{c}
%  0  \\
%  1  \\
%\end{array} \; \right) $.

The two-component wavefunctions are not orthogonal and normalized.
 This is not surprising since a non-unitary
transformation has been performed on the fermion field.
Eqs.~(\ref{tilde-u}) (\ref{tilde-v}) imply that the wavefunctions
of the fermion and the antifermion are different. It  can be used
to distinguish the fermion and the antifermion.

 We notice that for antifermions, ${\rm\bf p}=0$
is a singular point. However, the singularity cancels in the final
expression.

The free propagator of the fermion in the two-component form
$\widetilde G (p)$  can be read from the Lagrangian:
\begin{equation}\label{Gv}
\widetilde G (p) \;=\; i\;\frac{E+m}{p^2-m^2 + i \epsilon}\;.
\end{equation}
Two poles correspond to the contributions from the fermion and the
antifermion. In fact, this propagator can be decomposed as:
\begin{equation}\label{Gv-dec}
\widetilde G (p) \;=\; {i\over 2 \omega }
 \; \left( \;
 {\;\sum\limits_s \,\tilde{u}_s(\tilde p)\,
 \bar{\tilde u}_s(\tilde p)\;
 \over E - \omega + i \epsilon }
 \;+\;
 {\;\sum\limits_s \,\tilde {v}_s(\tilde p)\,
 \bar{\tilde v}_s(\tilde p)\;
 \over E + \omega - i \epsilon }
  \;\right) \;,
\end{equation}
where $ \omega \equiv \sqrt{ {\rm \bf p}^2 +m^2}$,
   $\tilde p \equiv (\omega, {\rm\bf p} ) $. This is just the
   result that we expect to have.

With the $\widetilde{G} (p) $ given in (\ref{Gv}), we can show the
following identities:
\begin{eqnarray}
\lim\limits_{p^2\to m^2}\; G(p)\;\mathrm{P}_{ + }
\;\widetilde{G}^{ - 1} (p)\;\widetilde u_s(p)\; &=&  \;u_s(p)\;,
 \label{uut} \\
 \lim\limits_{p^2\to m^2}\; G(-p)\;\mathrm{P}_{ + } \;\widetilde{G} ^{ - 1} (-p)\;
 \widetilde v_s(p)\; &=&
\; v_s(p)\;.
 \label{vvt}
\end{eqnarray}

Similar to Eqs.~(\ref{T-scat}) and (\ref{T-anni}), the $S-$matrix
element for the fermion scattering process in the two-component
form reads
\begin{eqnarray}
 \widetilde{T}_{s\,s'}(p'\to p + X ) \;& = &\;
 \lim\limits_{p^2,p'^2\to m^2}\; \bar{\tilde{u}}_s (p )
 \;\widetilde {G}^{ - 1} (p )\; \widetilde{G} (p ,p' ;\Gamma )\;\widetilde {G}^{ - 1}(p' )\;\tilde u_{s'}(p' )
 \;, \label{T'-scat}
\end{eqnarray}
and that for the annihilation process reads
\begin{eqnarray}
 \widetilde{T}_{s\,s'}(p'+ p \to X ) \;& = &\;
 \lim\limits_{p^2,p'^2\to m^2}\; \bar {\tilde{v}}_s (p )
 \;\widetilde{G}^{ - 1} (-p )\;\widetilde{G}(-p ,p' ;\Gamma )
 \;\widetilde{G}^{ - 1}(p' )\;\tilde u_{s'}(p' )
 \;, \label{T'-anni}
\end{eqnarray}
where $\tilde{G} (p ,p' ;\Gamma )$ is the Green's function
evaluated in the two-component form. The relation between the
Green's functions (\ref{ppp}) implies that
\begin{equation}\label{g-g'}
\tilde{G} (\,\pm\, p ,p' ;\Gamma \,) \;=\; \mathrm{P}_+\;G(\,\pm\,
p ,p' ;\Gamma\, )\;\mathrm{P}_+ \;.
\end{equation}
Imposing  this equation and Eqs.~(\ref{uut}) and (\ref{vvt}) on
Eqs.~(\ref{T'-scat}) and (\ref{T'-anni}), we find that
\begin{eqnarray}
\widetilde{T}_{s\,s'}(p' \to p+ X ) \;& = &\;{T}_{s\,s'}(p'\to p
+ X ) \label{t-t'-scat} \;,\\
\widetilde{T}_{s\,s'}(p'+ p \to X ) \;& = &\;{T}_{s\,s'}(p'+ p \to
X ) \label{t-t'-anni}\;.
\end{eqnarray}
Thus, from  the relations (\ref{ppp}), we have shown that the
$S-$matrix elements both for scattering processes and for
annihilation processes are the same in both formulations. This
demonstration can easily be generalized to the cases of
multi-fermionic external lines.

Obviously, the two-component form simplifies significantly the
$\gamma-$matrix algebra in the conventional four-component form.
Instead of the $4\times4$ $\gamma$ matrixes, only the $2\times 2 $
$\sigma$ matrixes are involved in the former case. This feature
makes it very useful in practical applications.  The calculations
for the $T-$matrix elements of the processes with multi-external
lines are very complicated. The helicity amplitude
method\cite{Xu:1986xb} and the direct amplitude
method\cite{Hagiwara:1988pp} have been proposed to simplify the
calculations. However, these methods are efficient only for
massless fermions. The two-component form presented here is
equally well both for massless fermions and for massive fermions.
 The perturbative calculations can easily be carried out in
the axial gauge $A\cdot V = 0$. In this gauge, the second sector
is a free term without any interaction. Hence all the
contributions to the $S-$matrix elements arise from the first
sector. The Lagrangian of this term reads:
\begin{eqnarray}
\widetilde{L}
 \;&=&\;
  \overline{\Psi}_+(x) \; \left( \,i \partial_0 \,-m\,\,+\,
  {\mathbf{D} \cdot \mbox{\boldmath $\sigma$} }\,
   \,\displaystyle{1 \over i\partial_0+m}\,
   {\mathbf{D} \cdot  \mbox{\boldmath $\sigma$} }  \,\right)\; \Psi_+(x)
 \;.
 \label{L-a0}
\end{eqnarray}
The Feynman rules can easily be obtained. The propagator of the
free fermion has been given in (\ref{Gv}).  Here we list the
Feynman rules of the interaction vertices as follows:

\vspace{-80pt} \hfill \\
\begin{center}
\SetScale{0.8}
\begin{tabular}{lc}
  %\hline
   \begin{picture}(130,80)(-20,40)
   % \SetWidth{0.7}
   %\Vertex(193.3,10){2} \Vertex(106.7,10){2} \Vertex(150,40){2}
    \SetColor{Blue}
    \Line(0,24)(100,24)
    \Line(0,26)(100,26)
    \Text(-20,20)[]{$p_2$}
    \Text(100,20)[]{$p_1$}
    \Text(37,70)[]{$k,\,i,\,a$}
    \SetColor{OliveGreen}
    \Line(27,22)(33,25)
    \Line(27,28)(33,25)
    \Line(67,22)(73,25)
    \Line(67,28)(73,25)
    \SetWidth{0.7}
    \Gluon(50,25)(50,75){5}{5}
     \SetWidth{0.5}
    \CCirc(50,25){3}{Red}{White}
    \SetColor{Green}
    \LongArrow(35,45)(35,60)
\end{picture}&
~~~~ $\displaystyle i\,g_s \,t^a \;\left(
{\;\frac{{\mbox{\boldmath $\sigma$} \cdot {\bf{p}}_1 }\;\sigma ^i
}{{E_1  + m}}\;  + \;\frac{\sigma ^i\;\mbox{\boldmath $\sigma$}
\cdot {\bf{p}}_2  }{{E_2  + m}}\;} \right)\;$
 \\
  %\hline
  \begin{picture}(130,80)(-20,35)
    \SetColor{Blue}
    \Line(0,24)(100,24)
    \Line(0,26)(100,26)
    \SetColor{OliveGreen}
    \Line(27,22)(33,25)
    \Line(27,28)(33,25)
    \Line(67,22)(73,25)
    \Line(67,28)(73,25)
    \SetWidth{0.7}
    \Gluon(50,25)(20,75){5}{5}
    \Gluon(50,25)(80,75){5}{5}
    \SetColor{Green}
    \LongArrow(23,40)(16,50)
    \LongArrow(75,40)(82,50)
     \SetWidth{0.5}
    \CCirc(50,25){3}{Red}{Red}
    \Text(-20,20)[]{$p_2$}
    \Text(100,20)[]{$p_1$}
    \Text(70,73)[]{$k_1,\,i,\,a$}
    \Text(10,73)[]{$k_2,\,j,\,b$}
    \end{picture} & ~~~~~~~
 $\displaystyle -i\,g_s^2 \;\left( {t^a t^b \;\frac{{\sigma ^i \;\sigma ^j }}{{E_1 +
k_1^0  + m}}\;  + \; t^b t^a \;\frac{{\sigma ^j \;\sigma ^i
}}{{E_1  + k_2^0  + m}}\;} \right)\;$ \\
  %\hline
\end{tabular}
\end{center}
\vspace{0pt} \hfill \\
With these rules and (\ref{tilde-u}), (\ref{tilde-v}), and
(\ref{Gv}), perturbation calculations can be carried out in the
two-component form. Since the propagator of fermion given in
eq.~(\ref{Gv}) is a scalar and the most generic interaction vertex
can be written as $a+b_i \sigma _i $, whose product can be easily
dealt with,  we expect that the numeric perturbation calculations
can be significantly simplified.

Another possible application is to solve the bound state problem.
In the two-component form, the simpler spinor structure of the
wavefunction and the equations makes bound state problems  easier
to be solved.

The two-component form might also be used to formulate chiral
fermion on Lattice to overcome the doubling problem\cite{lattice}.

In conclusion, we show that the relativistic massive fermions in
gauge theories can be described with a two-component form. In this
form, the fermion and the anti-fermion are uniformly expressed as
the two-component field instead of the usual Dirac four-component
field. We have shown that the $S-$matrix elements in this form are
exactly the same as that in the conventional four-component form.
The two-component form simplifies significantly the
$\gamma$-matrix algebra in the four-component form. It will be
very useful in calculating the high energy processes and in
solving the bound state problems. We also expect that it is useful
in simulating the chiral fermions on the lattice.

\noindent{\bf Acknowledgment:} The author wishes to thank X.~D.~Ji
for reading this draft carefully.  This work is partially
supported by the National Natural Science Foundation of China
(NSFC).


\begin{references}
\bibitem{dirac}
P.~A.~M.~Dirac, Proc. Roy. Soc. (London), {\bf A117}, 610 (1928);
{\it ibid } {\bf A118}, 351 (1928).

\bibitem{HQET}
%\bibitem{Isgur:vq}
N.~Isgur and M.~B.~Wise,
%``Weak Decays Of Heavy Mesons In The Static Quark Approximation,''
Phys.\ Lett.\ B {\bf 232}, 113 (1989);
%%CITATION = PHLTA,B232,113;%%
%\cite{Isgur:ed}
%\bibitem{Isgur:ed}
%``Weak Transition Form-Factors Between Heavy Mesons,''
Phys.\ Lett.\ B {\bf 237}, 527 (1990);
%%CITATION = PHLTA,B237,527;%%
H.~D.~Politzer and M.~B.~Wise,
%%We did not have this you might want to check it.%%
Phys.\ Lett.\  {\bf 206B}, 681 (1988);
%%CITATION = PHLTA,206B,681;%%
%%We did not have this you might want to check it.%%
Phys.\ Lett.\  {\bf 208B}, 504 (1988);
%%CITATION = PHLTA,208B,504;%%
%\cite{Eichten:1989zv}
%\bibitem{Eichten:1989zv}
E.~Eichten and B.~Hill,
%``An Effective Field Theory For The Calculation Of Matrix Elements Involving Heavy Quarks,''
Phys.\ Lett.\ B {\bf 234}, 511 (1990);
%%CITATION = PHLTA,B234,511;%%
%\cite{Georgi:1990um}
%\bibitem{Georgi:1990um}
%H.~Georgi,
%``An Effective Field Theory For Heavy Quarks At Low-Energies,''
Phys.\ Lett.\ B {\bf 240}, 447 (1990).
%%CITATION = PHLTA,B240,447;%%
%\cite{Grinstein:1990mj}



\bibitem{NRQCD}
%\cite{Caswell:1985ui}
%\bibitem{Caswell:1985ui}
W.~E.~Caswell and G.~P.~Lepage,
%``Effective Lagrangians For Bound State Problems In QED, QCD, And Other Field Theories,''
Phys.\ Lett.\ B {\bf 167}, 437 (1986);
%%CITATION = PHLTA,B167,437;%%
%%We did not have this you might want to check it.%%
G.P. Lepage and B.A. Thacher, Nucl.\ Phys.\ Proc.\ Suppl.\  {\bf
4}, 199 (1988);
%%CITATION = NUPHZ,4,199;%%

%\cite{Chen:1993sx}
\bibitem{Chen:1993sx}
Y.~Q.~Chen,
%``On the reparametrization invariance in heavy quark effective theory,''
Phys.\ Lett.\ B {\bf 317}, 421 (1993).
%%CITATION = PHLTA,B317,421;%%

\bibitem{Grinstein:1990mj}
 B.~Grinstein,
%``The Static Quark Effective Theory,''
Nucl.\ Phys.\ B {\bf 339}, 253 (1990).


%\cite{Chen:2002yr}
\bibitem{Chen:2002yr}
Y.~Q.~Chen,
%``Renormalization in reparameterization invariance,''
Phys.\ Rev.\ D {\bf 69}, 096001 (2004).
%%CITATION = HEP-PH 0207333;%%
%\cite{Kleiss:1985yh}

%\cite{Xu:1986xb}
\bibitem{Xu:1986xb}
Z.~Xu, D.~H.~Zhang and L.~Chang,
%``Helicity Amplitudes For Multiple Bremsstrahlung In Massless Nonabelian Gauge
%Theories,''
Nucl.\ Phys.\ B {\bf 291}, 392 (1987).
%%CITATION = NUPHA,B291,392;%%



%\cite{Hagiwara:1988pp}
\bibitem{Hagiwara:1988pp}
K.~Hagiwara and D.~Zeppenfeld,
%``Amplitudes For Multiparton Processes Involving A Current At E+ E-, E+- P, And
%Hadron Colliders,''
Nucl.\ Phys.\ B {\bf 313}, 560 (1989).
%%CITATION = NUPHA,B313,560;%%




\bibitem{lattice}
Y.~Q.~Chen and X.~D.~Ji, in preparation.


\end{references}
\end{document}